\documentclass[twocolumn]{aastex631}

\newcommand{\Swift}{\emph{Swift}}

\bibpunct{(}{)}{;}{a}{}{,}
\usepackage{graphicx}
\usepackage{amsmath}
\usepackage{gensymb}
\usepackage{tabularx}
\usepackage{comment}
\usepackage{txfonts}
\usepackage{hyperref}
\hypersetup{
    colorlinks=true,
    linkcolor=blue,
    filecolor=magenta,      
    urlcolor=blue,
    citecolor=blue
}

\graphicspath{{./}{figures/}}
\received{}
\revised{}
\accepted{}
\submitjournal{ApJ}
\shorttitle{Hadronic processes at work in 5BZB J0630-2406}
\shortauthors{Fichet de Clairfontaine et al.}
\usepackage{array}
\usepackage{changepage}
\usepackage{ulem}

\RequirePackage{soul}
\RequirePackage{xcolor}

\def\xmm{{XMM-{\it Newton\/}}}

\def\nus{{\it NuSTAR}}
\newcommand{\target}{5BZB\,J0630$-$2406}

\begin{document}

\title{Hadronic processes at work in \target}

\email{gaetan.fichet-de-clairfontaine@physik.uni-wuerzburg.de}

\author[0000-0002-0786-7307]{Gaëtan Fichet de Clairfontaine}
\affil{Julius-Maximilians-Universität Würzburg, Fakultät für Physik und
Astronomie, Emil-Fischer-Str. 31, D-97074 Würzburg, Germany}
\author[0000-0002-3308-324X]{Sara Buson}
\affil{Julius-Maximilians-Universität Würzburg, Fakultät für Physik und
Astronomie, Emil-Fischer-Str. 31, D-97074 Würzburg, Germany}
\author[0000-0003-2497-6836]{Leonard Pfeiffer}
\affil{Julius-Maximilians-Universität Würzburg, Fakultät für Physik und
Astronomie, Emil-Fischer-Str. 31, D-97074 Würzburg, Germany}
\author[0000-0001-5544-0749]{Stefano Marchesi}
\affil{Dipartimento di Fisica e Astronomia (DIFA), Università di Bologna, via Gobetti 93/2, I-40129 Bologna, Italy}
\affil{Department of Physics and Astronomy, Clemson University, Kinard Lab of Physics, Clemson, SC 29634, USA}
\affil{INAF - Osservatorio di Astrofisica e Scienza dello Spazio di Bologna, Via Piero Gobetti, 93/3, 40129, Bologna, Italy}
\author[0000-0002-2515-1353]{Alessandra Azzollini}
\affil{Julius-Maximilians-Universität Würzburg, Fakultät für Physik und
Astronomie, Emil-Fischer-Str. 31, D-97074 Würzburg, Germany}
\author[0000-0003-0477-1614]{Vardan Baghmanyan}
\affil{Julius-Maximilians-Universität Würzburg, Fakultät für Physik und
Astronomie, Emil-Fischer-Str. 31, D-97074 Würzburg, Germany}
\author[0000-0002-8186-3793]{Andrea Tramacere}
\affil{Department of Astronomy, University of Geneva, Ch. d’Ècogia 16, Versoix, 1290, Switzerland}
\author[0000-0003-4704-680X]{Eleonora Barbano}
\affil{Julius-Maximilians-Universität Würzburg, Fakultät für Physik und
Astronomie, Emil-Fischer-Str. 31, D-97074 Würzburg, Germany}
\author[0000-0003-4519-4796]{Lenz Oswald}
\affil{Julius-Maximilians-Universität Würzburg, Fakultät für Physik und
Astronomie, Emil-Fischer-Str. 31, D-97074 Würzburg, Germany}

\begin{abstract}
  Recent observations are shedding light on the important role that active galactic nuclei (AGN) play in the production of high-energy neutrinos. 
  In this study, we focus on one object, \target, which is among the blazars recently proposed as associated with neutrino emission during the first $7$-yr IceCube observations.
  Modelling the quasi-simultaneous, broad-band spectral energy distribution, we explore various scenarios from purely leptonic to lepto-hadronic models, testing the inclusion of external photon fields. This theoretical study provides a complementary testing ground for the proposed neutrino-blazar association. Despite being historically classified as a BL Lac, our study shows that 5BZB J0630$-$2406 belongs to the relatively rare sub-class of high-power flat-spectrum radio quasars (FSRQs).
 Our results indicate that interactions between protons and external radiation fields can produce a neutrino flux that is within the reach of the IceCube detector. Furthermore, the spectral shape of the X-ray emission suggests the imprint of hadronic processes related to very energetic protons.
\end{abstract}

\keywords{BL Lacertae objects: individual: \target~ -- gamma rays: galaxies -- neutrinos -- radiation mechanisms: non-thermal}

\section{Introduction}
\label{sec: Introduction}

Active galactic nuclei (AGN) are among the most energetic and powerful objects in the universe. They are powered by a supermassive black hole (SMBH), in some of them a relativistic jet can be present and detected across the electromagnetic spectrum, from the radio to the very high energy (VHE) band. Efficient particle acceleration mechanisms such as magnetic reconnection \citep{Blandford_2017} and shock acceleration \citep{Lemoine_2019, Lemoine_2019_2} can occur in distinct regions of the jet. Standing and moving features observed in AGN jets \citep{Marshall_2002, Jorsatd_2013, Lister_2021}, also reproduced numerically as shocks \citep{Fromm_2016, Fichet_2021, Fichet_2022}, support the idea that the observed multi-wavelength (MWL) electromagnetic emission can be explained trough radiative cooling of those accelerated particles along the jet. While to date the overwhelming majority of AGN can be explained invoking the leptonic framework, from the theoretical point of view relativistic jets harbored in AGN may be capable of accelerating hadrons. Within such scenarios, high-energy (HE; TeV/PeV energies) neutrinos are a natural by-product. 

Located at the geographic South Pole, the IceCube observatory is the most sensitive high-energy ($\gtrsim$~TeV) neutrino detector currently operating. Since the beginning of its science operations in $2008$, it allows us to study the putative neutrino counterparts of astrophysical sources. Although no firm associations between individual high-energy IceCube events and cosmic sources have been established to date, several claims of associations with AGN have been made at different statistical levels, e.g. the blazar TXS\,0506$+$56, \cite{IceCube_2018a, IceCube_2018b}, which represent a subclass of AGN with the jet pointed directly at the observer.

Other approaches to studying such correlations exploit the time-integrated neutrino information over a given period of time, as in the case of observational evidence of neutrino emission in the direction of the Seyfert galaxy NGC~1068 \citep{IceCube_2022}, or through stacking analyses of populations of sources \citep[e.g.][]{IceCube2017_2LAC,Padovani_extreme_blazars:2016, Aartsen_2017, IceCube_AGNcores:2021,Plavin_2021,IceCat-1:2023}. 

A recent work reports evidence for a statistically significant correlation between blazars listed in the fifth data release of the Roma-BZCat catalog  \citep[5BZCat, see][]{Massaro_2015}, and a sample of IceCube hotspots, i.e. anisotropies in the distribution of IceCube events \citep{Buson:2022, Buson_erratum:2022}. The study is based on the $7$-yr IceCube southern sky map published by the IceCube collaboration and highlights $10$ objects as candidate HE neutrino emitters, i.e. PeVatron blazars.
Consistent findings are reported when expanding the investigation to the Northern Hemisphere with the latest sky map released by the IceCube collaboration \citep{Buson_2023}.
Investigations employing different analysis methodologies have provided mixed results. For instance, \cite{Bellenghi_2023} performed an independent analysis of the public IceCube dataset, confirming the association established by \citet{Buson:2022, Buson_erratum:2022} 
with the $7$-yr Southern Hemisphere dataset. 
However, when analyzing the $10$-year dataset, they did not observe a similar correlation. Such $10$-yr sky map appears overall different from the one published by IceCube collaboration with the same dataset \citep{IceCube10y:2020}, and the one used by \citep{Buson_2023} for the $10$-yr Northern Hemisphere analysis.
The discrepancies can be attributed to the different likelihood formalism and the coarseness of the detector response matrices employed, that lead to an overall worst sensitivity, as acknowledged in \cite{Bellenghi_2023}.

Given the unascertained electromagnetic / neutrino relation, the statistical analysis was based solely on the positions of the objects. No a-priori selection was applied to the blazar sample, neither based on the objects' classification nor on their electromagnetic properties. 

This paper aims to provide an initial characterization of the underlying physics of PeVatron blazars from the theoretical perspective \citep{Fichet_2023}. To this extent, we focus on one object for which broad, MWL simultaneous observations are available, namely \target~(a.k.a. TXS\,0628$-$240, WISE\,J063059.51$-$240646.2), while future works will address the physical properties of the sample \citep{azzollini_2023}. 

Although the redshift of the source is unknown due to the lack of emission lines in the optical spectrum, a lower limit has been established $z \geq 1.239$ \citep{Shaw_2013}, and absorption lines have been observed (Mg~\MakeUppercase{\romannumeral1}, Fe~\MakeUppercase{\romannumeral2}, Al~\MakeUppercase{\romannumeral2}) that indicate the presence of the host galaxy \citep{Rau_2012}. 

Theoretical modeling of the MWL emission of candidate neutrino emitters have shown the key role of external fields in the neutrino production \citep{Dermer:2014, Oikonomou_2019} and markers of hadronic processes imprinted in the observed electromagnetic spectral energy distribution\citep[SED,][]{Reimer2019, Cerrruti_2019, Gao_2019, Petropoulou_2020, Rodrigues_2021}. 
Here, we model the SED of the object of interest, exploiting simultaneous and quasi-simultaneous MWL observations. 

The paper is organized as follows. Sect.~\ref{sec:neutrino} introduces the neutrino observations that provided evidence of neutrino emission in the direction of \target. The Sect.~\ref{sec:data} displays the MWL observations and data reduction. The Sect.~\ref{sec:variability} discusses the temporal variability of this source. We present our approach to reproduce the observed SED with a one-zone leptonic and lepto-hadronic numerical models, with a complete description in Sect.~\ref{sec: Methods}. Then, in Sect.~\ref{sec: Results}, we present the results of our parameter space exploration in different scenarios, including scenarios where the X-ray band is dominated by either synchrotron radiation or by hadronic cascade processes related to secondaries particles. The characteristics of the source derived from the various scenarios will be detailed, as well as the expected neutrino event rates. In Sect.~\ref{sec: Discussion}, we discuss our findings and their implications in the context of recent neutrino - blazar associations. \\
Throughout the paper, all primed quantities are evaluated in the rest frame of the relativistic jet. We also assume a flat $\Lambda$CDM cosmology with $\rm{H}_{\rm 0} = 69.6~\rm{km}\cdot\rm{s}^{-1}\cdot\rm{Mpc}^{-1}$, $\Omega_{\rm 0} = 0.29$ and $\Omega_{\Lambda} = 0.71$. 

\section{Neutrino observations}
\label{sec:neutrino} 

\begin{figure}
    \centering
    \includegraphics[width = 0.8\linewidth]{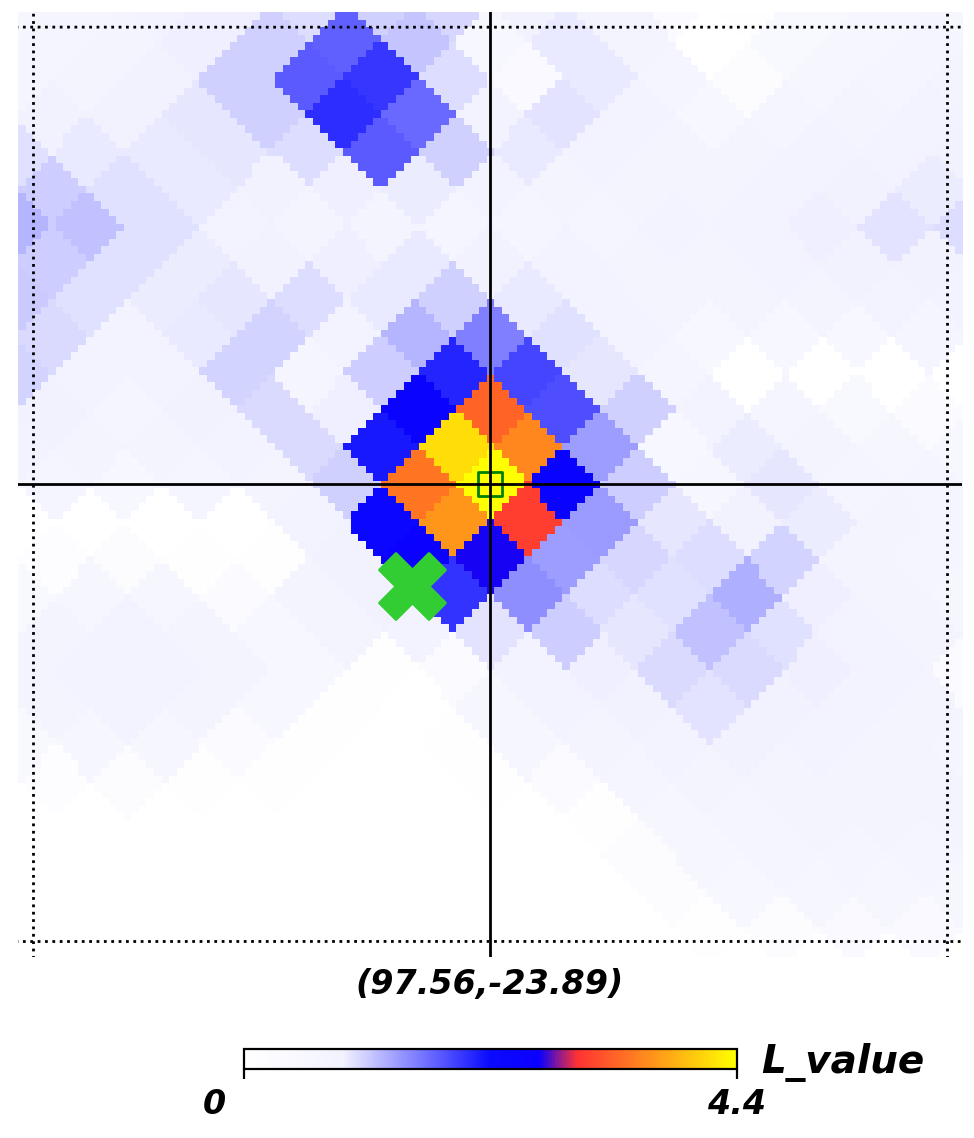}
    \caption{Cut-out region of the $7$-yr IceCube $L$-value map \citep{Aartsen_2017b} centered at the position of the hotspot IC\,J0630$-$2353, displayed in celestial coordinates. The position of the associated blazar \target~is highlighted by a green cross. The black lines indicate the $1$ degree $\times$ $1$ degree coordinate grid.}
    \label{fig: p_map}
\end{figure}

\target\ has been proposed as candidate counterpart of the IceCube hotspot IC\,J0630$-$2353 \citep{Buson:2022}. The study presented in \citet{Buson:2022} is based on a Southern Hemisphere analysis ($\delta < -5$) of the all-sky map that spans the IceCube observations from $2008$ to June $2015$. This $7$-yr sky map encodes the time-integrated information regarding long-term point-source neutrino emitters and is built using events of energy-proxy $\gtrsim100$~TeV. In the $7$-yr sky map, the direction of the IceCube hotspot IC\,J0630$-$2353 evidences an anisotropy in the spatial distribution of events (see Fig.~\ref{fig: p_map}) and, thus, may hint for the presence of astrophysical sources. The object is located at an angular separation of $0.28\degree$ from the hotspot IC\,J0630$-$2353, within the association radius of $0.55\degree$ derived for this dataset in \cite{Buson:2022}. The evidence for a spatial correlation between IC\,J0630$-$2353 and the blazar \target~suggests that this blazar may contribute to the observed anisotropy. 

\section{Multiwavelength observations and data analysis}
\label{sec:data}

 In the following, we provide a description of the MWL data collection and reduction. In Figure \ref{fig:MWL_panel}, we present the MWL light curves, from near-infrared (NIR) to $\gamma$ rays. The orange vertical line denotes the epoch of the simultaneous observations, used in the SED modeling presented in Sect. \ref{sec: Methods}.

\subsection{Radio observations}
\label{subsec:radio}

Archival radio data are available from the Academy of Sciences Radio Telescope (RATAN) \footnote{More details at \url{https://www.sao.ru/blcat/}}, the GaLactic and Extragalactic All-sky Murchison Widefield Array (GLEAM, \cite{Gleam:2017}) and the Australia Telescope 20-GHz (AT20G, \cite{AT20G:2011}) catalog. They are not used in the modeling and rather considered as upper limits for the fit. At such lower frequencies, the radio flux is likely originated from extended regions of the jet, as the electrons cool down via low-energy synchrotron over a long period of time while propagated through the jet. 

\subsection{Near-infrared observations} \label{NIR}
\label{subsec:NIR}

NIR, aperture photometry (JHK) observations were obtained by the Gamma-Ray Burst
Optical and Near-Infrared Detector (GROND) on $2010$--$09$--$28$~$08$:$06$  \citep[MJD $55467$,][]{Rau:2012} and $2014$--$10$--$19$~$05$:$29$ (MJD $56949$). The data were reduced and analyzed as explained in \citet{J0630:2016}.

\subsection{Optical observations}
\label{subsec:optical}

\begin{figure*}
    \centering
    \includegraphics[width=0.75\textwidth]{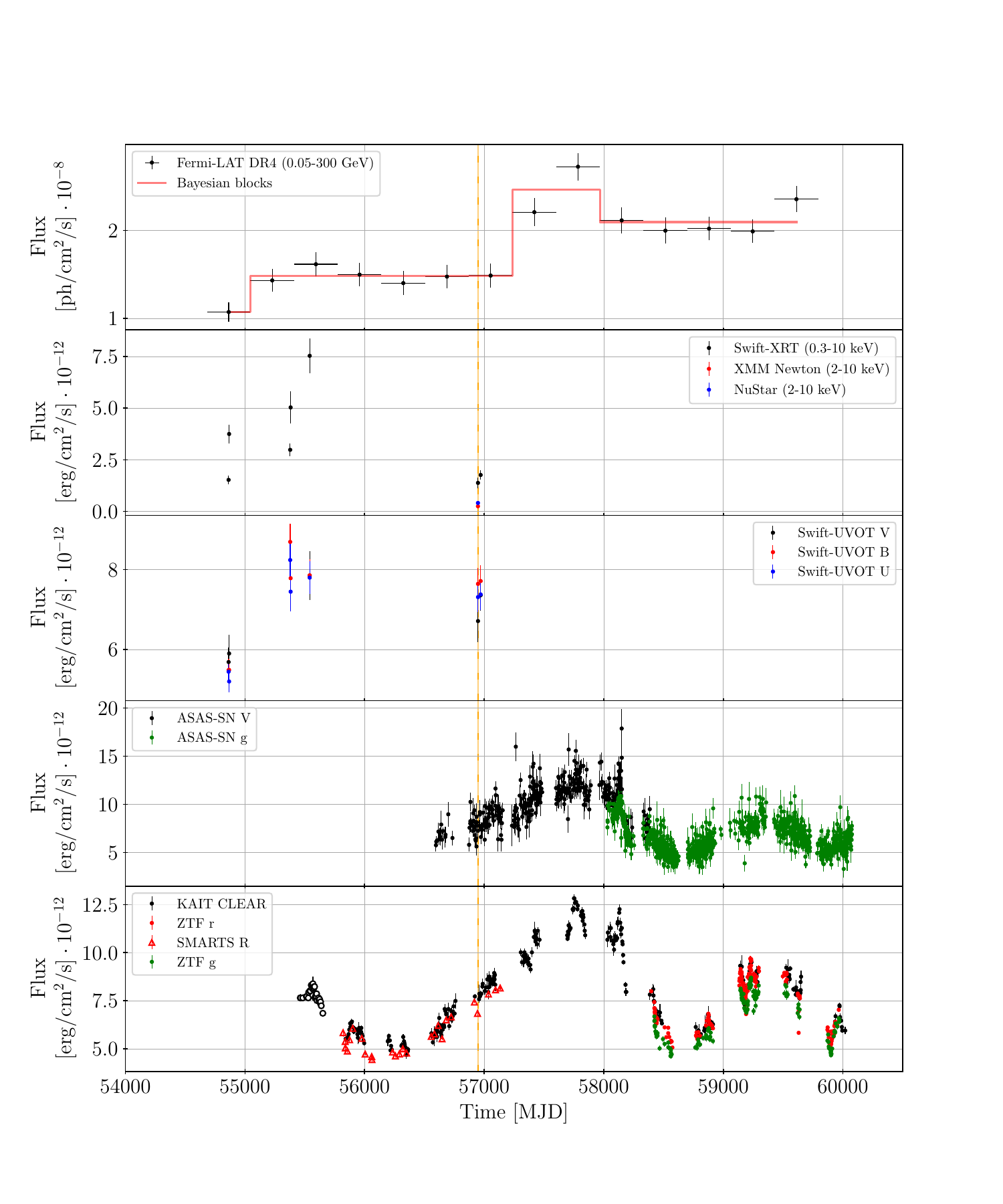}
    \caption{Multi-wavelength light curve of \target\ from  $\gamma$-rays, X-rays to the optical band. The top panel shows the $\gamma$-ray light curve with one-yr binning, available from the 4FGL-DR4 catalog. The second panel displays measurements from \Swift-XRT, XMM Newton and Nustar used for the modeling. The third panel shows values taken by \Swift-UVOT in the three optical filters V, B, and U. The fourth panel shows the V-band and g-band data collected by ASAS-SN and the bottom panel shows the KAIT $Clear$ band measurement, the SMARTS R-band and the ZTF r-band and g-band. The X-ray and optical light curves are corrected for galactic extinction. The orange line highlights the time of quasi-simultaneous data used in the SED analysis. GROND data also used for modeling but not included in this panel are discussed in section \ref{NIR}.}
    \label{fig:MWL_panel}
\end{figure*}

Over the past decade \target~has been monitored at optical wavelengths by several programs.
The Katzman Automatic Imaging Telescope  \citep[KAIT,][]{Filippenko:2001,Li:2003}  performed optical photometric observations in the $Clear$ band, which is close to the R band ($6410\,\textup{\AA}$), within the \emph{Fermi}-LAT AGN monitoring program \citep{Cohen:2014}. The data are calibrated to Landolt $R$ band, and corrected for the galactic reddening with $A_{\rm R}=0.138~\rm{mag}$, following \citet{Schlafly:2011}. The magnitudes are converted to flux units using the zero points \citep{Bessell:1998}. 
For earlier observations shown as black circles in Fig. \ref{fig:MWL_panel}, around MJD~$55500$, a manual comparison was made with calibrated data at corresponding points in time.

\target~has also been monitored by the All-Sky Automated Survey for Supernovae \citep[ASAS-SN,][]{Shappee:2014, Kochanek:2017} in the V-band and g-band. The magnitude corrections are applied using the reddening coefficient $\mathrm{E(B - V) = 0.054739}$ according to \citet{Schlafly:2011} with ratio of the extinction reddening $A_\lambda/\rm{E(B - V )}$ for each filter taken from \citet{Fitzpatrick:1999}. The conversion to flux units follows the same methodology as explained previously for KAIT.
The Zwicky Transient Facility (ZTF, \cite{Masci:2019}) monitored the source in the r-band and g-band, as well as the Small and Moderate Aperture Research Telescope System \citep[SMARTS,][]{Bonning:2012} in the R-band. The collected magnitudes are corrected for reddening and converted to flux units consistent to KAIT and ASAS-SN.

\Swift-UVOT observations are available in three optical filters (U, B, and V) between $2009$--$02$--$01$ (MJD $54863$) and $2014$--$11$--$10$ (MJD $56971$). The UVOT data above $\nu \geq 10^{15}~\rm{Hz}$ are affected by the $\rm{Ly}\alpha$ forest absorption and hence not used in this study.
We extracted source counts using apertures with a radius of $5$", while the background counts were obtained from an annulus region centered on the source position where no other sources were visible. To compute the magnitudes, we used the UVOTSOURCE tool in HEASOFT. Extinction corrections were applied following \citet{Schlafly:2011} and \citet{Fitzpatrick:1999} for each filter. Then, using the zero points derived from \cite{2011AIPC.1358..373B}, we subsequently converted the magnitudes to fluxes following the procedure described in \citet{2008MNRAS.383..627P}.

\subsection{X-ray observations}

\subsubsection{The Neil Gehrels \Swift\ observatory}

Overall,  \Swift-XRT \citep{2004ApJ...611.1005G} observed \target~eight times, including visits during the period of the IceCube observations. To analyze the \Swift-XRT data, the standard analysis tools provided by the HEASOFT\footnote{More details at \url{http://heasarc.nasa.gov/lheasoft/}} \texttt{V.6.31.1} software package were used.

The XRT observations were all in photon counting (PC) mode, and since the count rates were always below $0.5$ counts $\mathrm{s^{-1}}$, no pile-up correction was necessary. To extract the source events, a circular region with a radius of $45$" was chosen within the $0.3–10~\rm{keV}$ energy range. The background events were extracted from the annular region around the source position. Given the low photon statistics, we performed the spectral analysis with the Cash statistics method by rebinning the spectrum to ensure a minimum of one count per bin.

\subsubsection{\xmm\ and \textit{NuSTAR} data reduction}
\label{susec:nustar_xmm}

\target~ was targeted quasi-simultaneously by \xmm\ (obsID: \texttt{0740820401}; exposure: $9~\rm{ks}$) and \nus\ (obsID: \texttt{60001140002}; exposure: $66~\rm{ks}$) on October $17$--$18$, $2014$.
We retrieved the \xmm~ pn source and background spectra and associated matrices from the 4XMM catalog\footnote{More details at \url{http://xmm-catalog.irap.omp.eu/source/207408204010001}.} \citep{Webb_2020}. The source spectrum has been binned with $1$ count per bin, to avoid having empty bins that can affect the spectral fit.
To generate the \textit{NuSTAR} spectra, we followed the standard data reduction procedure. Specifically, the data have been processed using the \textit{NuSTAR} Data Analysis Software (NUSTARDAS) version \texttt{2.1.1}. The raw event files are calibrated by the \textit{nupipeline} script, using the response file from the Calibration Database (CALDB) version \texttt{20210202}. The source and background spectra are extracted from a $45^{\prime\prime}$ ($\approx50\%$ of the encircled energy fraction at $10~\rm{keV}$) circular region, centered at the optical position of the source and in a nearby ($\sim$3$^\prime$ separation) region which was visually inspected to avoid any possible contamination, respectively. 
Using \textit{nuproducts} scripts, we then generated source and background spectra files, along with the corresponding ARF and RMF files. Finally, the \textit{NuSTAR} spectra are grouped with $1$ count per bin, using \textit{grppha}. This procedure has been performed on both the \textit{NuSTAR} focal plane modules, FPMA and FPMB.

\subsection{\xmm~ and \nus~ spectral fitting results}
\label{susec:x-ray_fit}
We fitted the \xmm~ and \nus~spectra of our target using the \texttt{XSPEC} \citep{Arnaud_1996} software, version \texttt{12.12.1}. We fixed the metal abundance to Solar metallicity using the abundances from \cite{Wilms_2000}, while the photoelectric cross-sections for all absorption components are those derived by \cite{Verner_1996}. The Galactic absorption column density is fixed to $\rm{N}_{\rm H,gal} = 7.5 \times 10^{20}~\rm{cm}^{-2}$ \citep{Kalberla_2005}. To maximize the spectral statistic, we analyze the data with the Cash statistic \citep{Cash_1979}, which uses a Poisson likelihood function and is hence most suitable for low numbers of counts per bin. As mentioned in the previous section, we bin the spectra with $1$ count per bin. The \xmm~ pn spectrum is fitted in the $0.3$--$10~\rm{keV}$ band, while the two \nus~ FPMA and FPMB spectra are fit in the $3$--$70~\rm{keV}$ band.

Following the standard procedure for fitting X-ray spectra of blazars, we first fit our data with a simple power-law model. We also add a cross-normalization constant to take into account the fact that the \nus~ observation is significantly longer than the \xmm~ one, and to model systematic cross-instrument offsets in flux. 
Finally, we include a column density at the redshift of the source, $N_{\rm H,z,ISM}$, assuming a redshift of $1.239$, to account for a possible contribution of the interstellar medium (ISM) to the absorption of X-ray photons. We measure a best-fit X-ray photon index $\Gamma_{\rm X}=3.03\pm0.12$, with an ISM column density $N_{\rm H,z,ISM}=3.0_{-1.0}^{+1.1} \times 10^{21}~\rm{cm}^{-2}$, and a \xmm--\nus~cross-normalization $\rm{C}_{\rm XMM-NuS}=1.90_{-0.34}^{+0.41}$. The best-fit statistic for this fit is $\rm{Cstat/d.o.f.} = 1377.8/1408$.

We then fit the data with a log-parabola model, which has been shown to accurately describe the X-ray spectral shape of blazars \citep[see, e.g.,][]{Bhatta_2018}. In particular, \citet{Middei_2022} analyzed the \nus~ spectra of $126$ blazars and found that in some cases a log-parabola model provides a more statistically accurate description of a blazar X-ray spectrum. The log-parabola model is described by the following equation,

\begin{equation}
\dfrac{\rm{d}N}{\rm{d}E}= K \left ( \frac{E}{E_{0}} \right )^{-\left ({\alpha -\beta \log\left (E/E_{0}  \right )}  \right )},
\end{equation}
where $E_{0}$ is the reference energy, which we fix as $5~\rm{keV}$ (in the observer frame) following the results of previous works \citep[e.g.,][]{Massaro_2004,Balokovic_2016,Middei_2022}, while $\alpha$ and $\beta$ are the photon index and the curvature parameter, respectively \citep[e.g.,][]{Massaro_2004}, and $K$ is the model normalization. 

The log-parabola component best-fit parameters are $\alpha=2.71_{-0.17}^{+0.19}$, $\beta=-0.42_{-0.20}^{+0.21}$, and $N_{\rm H,z,ISM} = 5.5_{-1.7}^{+1.7} \times 10^{21}\rm{cm}^{-2}$, while the \xmm--\nus\ cross-normalization is $\rm{C}_{\rm XMM-NuS}=1.66_{-0.32}^{+0.39}$. The best-fit statistic for this fit is $\chi^2/\rm{d.o.f.} = 1367.4/1407$.
The difference in $\rm{Cstat}$ between this model and the simple power-law one is $\Delta \rm{Cstat} = 1377.8 - 1367.4 = 10.4$. Since the log-parabola model has $1$ degree of freedom less than the simple power-law one, the log-parabola model is statistically preferred\footnote{Typically, a $\Delta \rm{Cstat} \geq 2.71$ ($90$\,\% confidence level) is the criterion used to infer when an additional free parameter/spectral component is statistically required in an X-ray spectral fit \citep[see, e.g.,][]{Tozzi_2006,Brightman_2014,Marchesi_2016}.
} at the $\sim99.9$\,\% confidence level (i.e., with a $\gtrsim3.2\,\sigma$ significance).

Finally, we performed a fit with a broken power-law model, which is also commonly used to describe the X-ray spectra of blazars. The parameterization of the broken power-law is,
\begin{equation}
   \frac{\rm{d}N}{\rm{d}E}=
\begin{cases}
   K E^{-\Gamma_{1}}& \text{if } E\leq E_{\rm b}\,,\\
   K E_{\rm b}^{\Gamma_{2}-\Gamma_{1}}(E/1\,{\rm keV})^{-\Gamma_{2}})              & \text{otherwise.}
\end{cases}
\end{equation}

Here, $\Gamma_{1}$ and $\Gamma_{2}$  are the low and high-energy photon indexes, $K$ is the normalization parameter, and $E_{\rm b}$ is the (rest-frame) energy of the break. The best-fit parameters are $\Gamma_{1}=3.19_{-0.16}^{+0.17}$, $\Gamma_{2}=2.57_{-0.26}^{+0.24}$, $E_{\rm b}=6.3_{-2.2}^{+3.1}~\rm{keV}$ and $N_{\rm H,z,ISM} = 4.2_{-1.3}^{+1.3} \times 10^{21}~\rm{cm}^{-2}$. The \xmm--\nus\ cross-normalization is $\rm{C}_{\rm XMM-NuS}=1.61_{-0.37}^{+0.55}$ The best-fit statistic for this fit is $\rm{Cstat/d.o.f.} = 1365.5/1406$. The difference in $\rm{Cstat}$ between this model and the simple power-law one is $\Delta \rm{Cstat} = 1377.8 - 1365.5 = 12.3$. Since the  broken power-law model has $2$ fewer degree of freedom than the simple power-law one, the broken power-law model is statistically preferred at the $>99.8$\,\% confidence level (i.e., at a $\gtrsim 3.1~\sigma$ significance).  

\begin{figure}[!h]
    \centering
    \includegraphics[width=.44\textwidth]{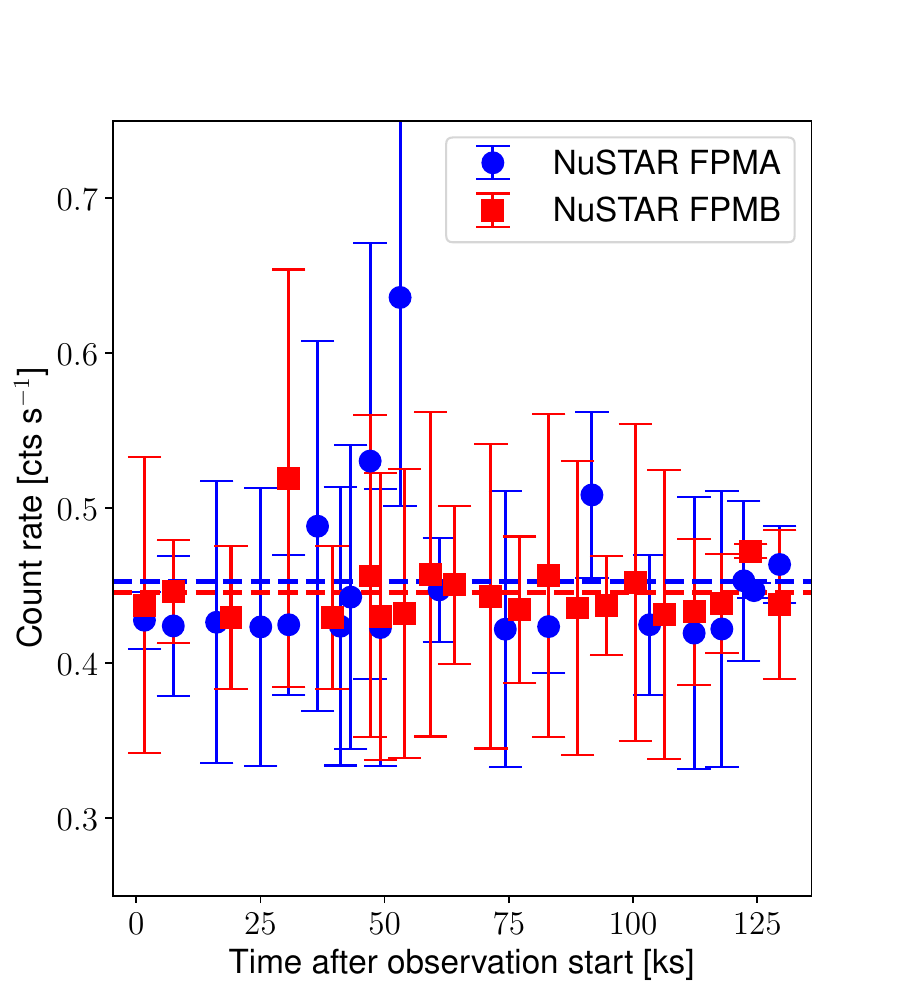}
    \caption{\target~\nus ~FPMA (blue circles) and FPMB (red squares) light curves. Both datasets are fit with a constant, shown with a blue (red) dashed line. No trend is observed in the light curves, suggesting no significant intra-observation variability in the \nus ~data.
    }
    \label{fig:light_curve_nustar}
\end{figure}

In summary, a two-component model is statistically favored with respect to a simple power-law one, while from a statistical point of view the log-parabola model and the broken power-law one are fully consistent. Differently from a previous study \citep{J0630:2016}, our more sensitive analysis supports the presence of a break in the X-ray spectrum. We report in Fig.~\ref{fig:light_curve_nustar} the source light curves from both the \nus~ cameras and, as it can be seen, no clear variability trend is observed within the \nus~ observation. More in detail, we fit the two light curves with a constant and obtain best-fit count rates $r_{\rm FPMA}=0.45~\rm{cts}\cdot \rm{s}^{-1}$ and $r_{\rm FPMB}=0.44~\rm{cts}\cdot \rm{s}^{-1}$, with reduced chi square ($\chi^2/\rm{d.o.f.})_{\rm FPMA}=9.49/20$ and  ($\chi^2/\rm{d.o.f.})_{\rm FPMB}=0.86/19$.
We also note that the \xmm--\nus~ cross-normalization values we measured in the log-parabola and broken power-law fits, while slightly higher than expected in quasi-simultaneous observations, have been observed in other \xmm--\nus~ quasi-simultaneous observations \citep[e.g.,][]{Marchesi_2019} and may be explained by a simple cross-instrument calibration offset.  

\subsection{Fermi-Large Area Telescope observations} \label{subsec:lat}

The study utilized gamma-ray data taken from \cite{J0630:2016}, and obtained from \emph{Fermi}-LAT observations spanning the period between August $4$, $2008$, and January $31$, $2015$ within the energy range of $100$ MeV to $500$ GeV. This time period is contemporaneous to the simultaneous X-ray data introduced in the previous section. 

\section{Temporal Variability}
\label{sec:variability}

The long-term optical light curves presented in Fig. \ref{fig:MWL_panel} highlight a similar flux variability pattern over the monitored bands.
According to the 4FGL-DR4 catalog \citep{Fermi_DR4}, the source is variable at $\gamma$-rays on $\gtrsim$\;year-timescales. To characterize  statistically-significant variations in the $\gamma$-ray light curve, we adopt the Bayesian algorithm available in Astropy\footnote{More details at \url{http://docs.astropy.org/en/stable/api/astropy.stats.bayesian_blocks.html}} \citep{2013A&A...558A..33A}. To determine the optimal value of the prior for the number of blocks, we use the empirical relation evaluated in \citet{Scargle:2013} for the probability to falsely report a detection of a change point, setting it to $0.05$. Applying this approach to the yearly binned light curve, the first $\sim7$-yr of LAT observations, up to $\sim$~MJD~$57250$, are overall consistent with a steady state, indicating that if variability is present at these frequencies it may be below the sensitivity of the LAT data. During the more recent LAT observations, since $\sim$~MJD~$57250$, the source is undergoing a long-term enhanced state. 

Optical evidence of year-long modulations of the flux, on timescales of $\sim3/4~\rm{yr}$. This is consistent with the  behavior traced by the $>15$-yr $\gamma$-ray light curve, including a major flux enhancement observed around MJD $57800$. The sparse X-ray/UV data mimic the overall variability pattern at other frequencies. The high-cadence KAIT monitoring, with sampling as short as a $\sim3~\rm{days}$ cadence, evidences statistically significant ($\gtrsim6\sigma$) changes between consecutive observations \citep{J0630:2016}. This suggests that at least some of the jet's emission arises in compact $R \lesssim 10^{16}~\rm{cm}$ regions, in the observer's frame. 

\section{Numerical SED modeling set up}
\label{sec: Methods}

\subsection{Building the quasi-simultaneous broadband SED}

In this work, we are mostly interested in the modeling of the MWL SED during the time span of the IceCube observations, i.e. $2008$-$2015$. 
Within this time range, MWL contemporaneous observations of \target\ were collected with the gamma-ray burst Optical/Near-Infrared Detector (GROND) instrument at the $2.2~\rm{m}$ MPG telescope at the ESO La Silla Observatory \citep{Greiner_2008}, the \Swift\ Niel Gehrels Observatory, XMM-Newton and NuSTAR \citep{Harrison_2013} satellites. They were performed around October $17$, $2014$ , i.e. $\sim~\rm{MJD}~56948$, close to the time period of the $7~\rm{yr}$ neutrino observations. These simultaneous ($\lesssim~1~\rm{day}$) multiwavelength data constrain the X-ray part of the synchrotron component. To constrain the second hump of the SED, we employ contemporaneous observations carried out by the $Fermi$-LAT in the MeV-GeV range. As discussed in Sec. \ref{sec:variability}, the source of interest is characterized by long-term variability in the $Fermi$-LAT band, and no significant variations in the flux are observed during the first $7$-yr of monitoring. Therefore, for the SED modeling we employ the LAT spectrum integrated over the first $7$ years of observations (from August $04$, $2008$ to August $15$, $2015$, i.e. MJD $54682-57250$) available from the literature \citep{J0630:2016}. We exclude the UVOT data above $\nu \geq 10^{15}~\rm{Hz}$ as they are affected by the $\rm{Ly}\alpha$ forest absorption. Fig. \ref{fig: purely leptonic SED} and Fig. \ref{fig: lepto hadronic SED 1} show the contemporaneous broadband SED built with these data (black points). Further archival radio, optical, and near-IR observations are displayed for comparison (gray points), and not included in the SED modeling.
The red downward triangle represents a limit on accretion disk luminosity from \citet{Ghisellini_2012}.

\subsection{One zone model}
\label{subsubsec: One zone model}

We explore different scenarios to describe the MWL emission of \target\ using the time-dependent code AM$^{\rm 3}$ \citep{Gao_2017}. This code is able to solve the system of coupled differential equations that describe the transport of particles through the relativistic jet. Relativistic electrons and protons are assumed to be accelerated initially and injected into a single zone where they can radiate their energy and interact with external photon fields. For more details on the implementation of the synchrotron, inverse Compton scattering, photo-photon pair annihilation, hadronic processes (i.e., Bethe-Heitler pair production and photo-pion production), we refer the reader to \cite{Gao_2017}. We also note that the synchrotron self-absorption opacity of the blob is treated only for the electrons (we expect a negligible contribution from the proton population in the radio band).

\begin{figure}
    \centering
    \includegraphics[width = 0.5\textwidth]{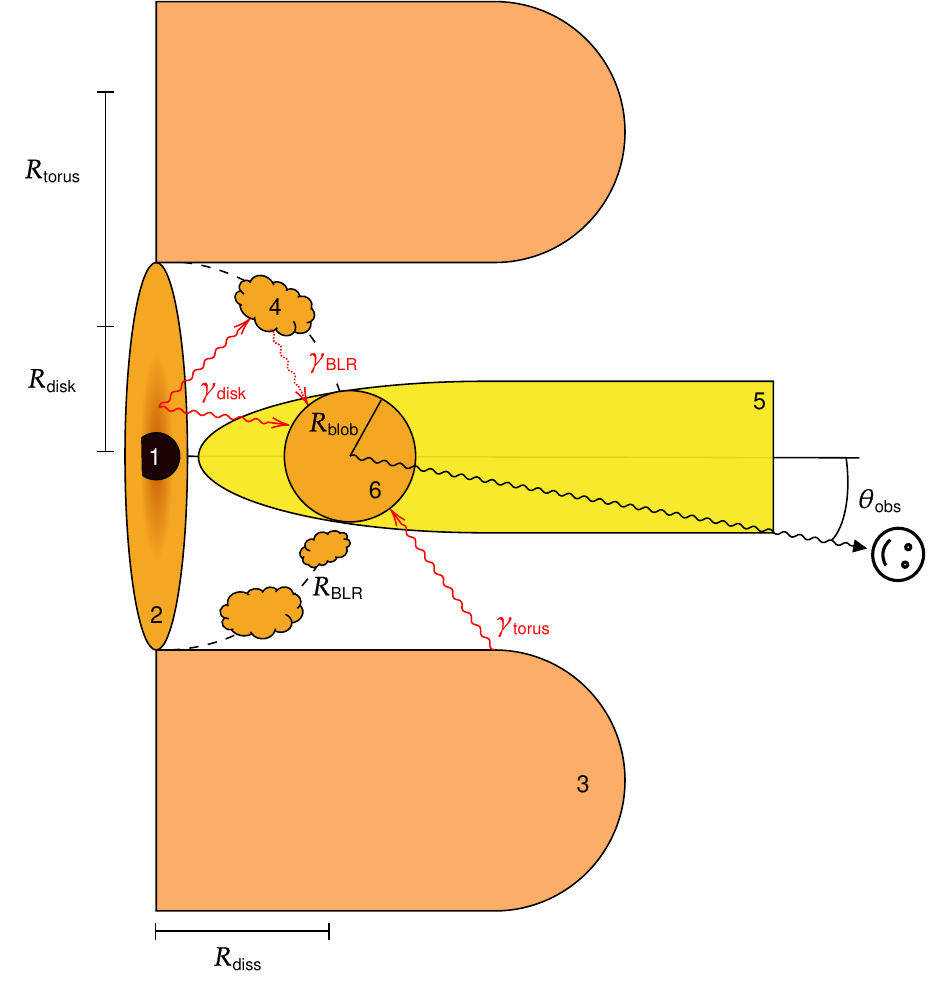}
    \caption{
    Schematic view of the model. 1: Supermassive black hole, 2: Accretion disk defined by a luminosity $L_{\rm disk}$ and a temperature $T_{\rm disk}$, 3: Dust torus defined by a luminosity $L_{\rm torus}$ and a temperature $T_{\rm torus}$, 4: Broad line region (BLR) that scattered direct emission, 5: Relativistic jet (not modeled here) and 6: Moving emission region (“blob”) situated at a distance $R_{\rm diss}$ from the black hole. Direct (indirect) external photon fields are represented by continuous (dashed) arrows. Scheme not to scale.}
    \label{fig: model geometry}
\end{figure}

As represented in the Fig.~\ref{fig: model geometry}, the emission region is modeled as a sphere of radius $R'_{\rm b}$ that moves at relativistic speed with a Lorentz factor of $\Gamma_{\rm b}$. From Earth, the blob is observed with an observation angle $\theta_{\rm obs} = 1/\Gamma_{\rm b}$, with quantities Doppler-shifted according to the Doppler factor $\delta_{\rm D} = \Gamma_{\rm b}$.
A magnetic field with a strength of $B'$ is assumed to be present in the blob and isotropic, i.e. with no preferential direction.
The blob is located at a distance of $R_{\rm diss}$ from the supermassive black hole, also known as the dissipation radius, considered static here. Since the energy density of external photon fields varies with $R_{\rm diss}$, this parameter is a crucial aspect of our model. 

In our model, relativistic electrons and/or protons are injected into the blob. For the electrons, we use a broken power-law distribution given by,

\begin{equation}
\dfrac{\rm{d}N_{\rm e}}{\rm{d}\gamma'_{\rm e}} \propto
\begin{cases}
{\gamma'_{\rm e}}^{-p_{\rm e, 1}} & \text{for } \gamma'_{\rm e, min} < \gamma'_{\rm e} \leq \gamma_{\rm e, brk} \,, \\
{\gamma'_{\rm e}}^{-p_{\rm e, 2}} \cdot {\gamma'_{\rm e, brk}}^{p_{\rm e, 2} - p_{\rm e, 1}} & \text{for } \gamma'_{\rm e, brk} < \gamma'_{\rm e} < \gamma'_{\rm e, max} \,,
\end{cases}
\end{equation}

where $\gamma'_{\rm e, min}$ and $\gamma'_{\rm e, max}$ are the cut-off values, and $\gamma'_{\rm e, brk}$ is the break energy. The choice of this distribution is supported by a pre-fit analysis on the observed SED that it used for initial guesses only (as it can be seen in \cite{Ghisellini_2012}). For protons, we assume a simple power-law distribution,

\begin{equation}
\dfrac{\rm{d}N_{\rm p}}{\rm{d}\gamma'_{\rm p}} \propto {\gamma'_{\rm p}}^{-p_{\rm p}}~~~\text{for } \gamma'_{\rm p, min} < \gamma'_{\rm p} < \gamma'_{\rm p, max} \,.
\end{equation}

The parameters $N_{\rm e}$ and $N_{\rm p}$ are the normalization factors. We ensure that the constraints on $\gamma'_{\rm e, brk}$ and $\gamma'_{\rm e/p, max}$ are compatible by respectively equating the synchrotron cooling timescale $\tau_{\rm syn}$ with the adiabatic timescale $\tau_{\rm ad} = 2 R'_{\rm b} / c$ and the acceleration timescale, $\tau_{\rm acc}$\footnote{Here, we define the acceleration timescale as,
\[
    \tau_{\rm acc} = \frac{1}{\eta} \frac{m_{\rm e, p}}{eB} \gamma_{\rm e, p} \,, 
\]
where we assume $\eta = 0.1$ \citep{Cerruti_2015}.}
with the shortest cooling timescale. For $\gamma'_{\rm e/p, max}$, we also check that the Hillas criterion is satisfied.

Given the expected large redshift, we incorporate the effect of absorption by the extragalactic background light (EBL) in our model. To achieve this, we use the Python library \texttt{ebltable}\footnote{Available at \url{https://github.com/me-manu/ebltable}}, which is based on the model presented by \cite{Dominguez_2011}.

The presence of a luminous accretion disk and radiation fields has been suggested by previous literature studies \citep{Ghisellini_2012,Padovani_2012}. External radiation field radiation can either interact directly with the source in the jet or be re-processed by a broad line region (BLR). A dust torus is also present in this model, as its presence has been considered in numerous previous studies \citep{Finke_2016, Murase_2014, Oikonomou_2021}. An upper limit on the disk luminosity has been derived in \cite{Ghisellini_2012},

\begin{equation}
    L_{\rm disk} \leq 5.5 \times 10^{45}~\rm{erg} \cdot \rm{s}^{-1} \,.
\end{equation}

It was obtained assuming an empirical relation between the BLR and gamma-ray luminosities ($L_{\rm BLR} \sim 4 \times L_{\rm \gamma}^{0.93}$).

Therefore, the model presented accounts for both external radiation fields originated either from the accretion disk and the dust torus, modeling them in the observer frame as a single temperature black body emission \citep{Dermer_2009} for simplicity, as visible in \cite{Rodrigues_2019}. This choice is also motivated by the lack of optical lines, as it implies a higher uncertainty on the black hole mass and the disk type. 

Following \cite{Ghisellini_2017}, we assume in this study that the luminosity and the size of the dust torus can be derived from the disk luminosity and the dust torus temperature, 

\begin{align}
    L_{\rm torus} &= L_{\rm disk} / 2\,, \\ 
    R_{\rm torus} &= 3.5 \times 10^{18}~\left(\dfrac{L_{\rm disk}}{10^{45}~\rm{erg.s}^{-1}}\right)^{1/2} \cdot \left(\dfrac{T_{\rm torus}}{10^3~\rm{K}}\right)^{-2.6} \,.
 \end{align}

where  $L_{\rm disk}$, $T_{\rm disk}$, and $T_{\rm torus}$ are free parameters in the model. 
Our model assumes that the emission from the accretion disk is generated by a single-temperature black-body that interacts with the source jet through a Doppler de-boosting effect, which is proportional to $\Gamma_{\rm b}^2$. In fact, we consider a scenario in which a single-temperature accretion disk emits an isotropic radiation “behind” the blob, as the jet is pointing at the observer. The emission of the dust torus is boosted in the blob frame and also consider isotropic. Additionally, the emission can be scattered by the BLR, which we assume is modeled as a thin shell located at a distance $R_{\rm BLR}$ from the central engine and radiating the luminosity $L_{\rm BLR}$. Both $L_{\rm BLR}$ and $R_{\rm BLR}$ are derived following \citet{Ghisellini_2008},

\begin{align}
    L_{\rm BLR} &= 10^{-1} \cdot L_{\rm disk}\,, \\ 
    R_{\rm BLR} &= 10^{17}~\rm{cm} \cdot \left(\dfrac{L_{\rm disk}}{10^{45}~\rm{erg.s}^{-1}} \right)^{1/2} \,.
\end{align}

We assume that the BLR reprocesses $10\%$ of the disk emission isotropically in the rest frame of the black hole \citep{Sbarrato_2012}. In fact, most of the reprocessed flux is emitted in emission lines (e.g. Lyman $\alpha$) and not as a thermal continuum (which accounts for $1\%$ \citep{Blandford_1995, Murase_2014}). In fact, we consider that most of them will lie on the disk emission range (Lyman $\alpha$ is situated at $\sim 10^{15}~\rm{Hz}$) and are outshined by the non-thermal continuum from the jet \citep{Rodrigues_2021}. \\
In the blob frame, this radiation field is enhanced by a factor $\delta^2$. This Doppler factor depends on $\Gamma_{\rm b}$ but also on the dissipation radius $R_{\rm diss}$, so that the energy density perceived by the blob will be lower outside the BLR radius, according to scaling factors calculated in \cite{Ghisellini_2009}. 

To explore the parameter space, we develop a parameter search algorithm which is parallelized, with initial and mutation guesses stored on the first and successive central processing units (CPU), respectively. Gaussian noises are applied to each parameter as mutations. We use the \texttt{least\_squares} function from the \texttt{scipy.optimize}\footnote{Additional information on the optimization algorithms used in this work can be found at \url{https://docs.scipy.org/doc/scipy/reference/optimize.html}.} library to optimize the parameters. For each parameter set and CPU, the algorithm calls the AM$^3$ code, which returns the simulated SED. The optimization is based on minimizing the residuals. Finally, we select the parameter sets associated with the best $\chi^2/{\rm d.o.f.}$ value. Here, the $\rm{d.o.f.}$ term refers to the number of degrees of freedom. If the $\chi^2/{\rm d.o.f.}$ value is not acceptable, we use this latest set of parameters found as input for the next generation, with the other solutions being mutations of this one. The process stops when the $\chi^2/{\rm d.o.f.}$ value is sufficiently low, i.e., $\chi^2/{\rm d.o.f.} \leq 2$. More details on the parameter space research, and on the evaluation of initial guesses, are show in Appendix~\ref{sec: Parameter space research}. 

\vspace{1cm}

\section{SED modeling Results}
\label{sec: Results}

\subsection{Purely leptonic solution}
\label{subsec: Purely leptonic solution}

\begin{figure*}
    \centering
    \includegraphics[width = \textwidth]{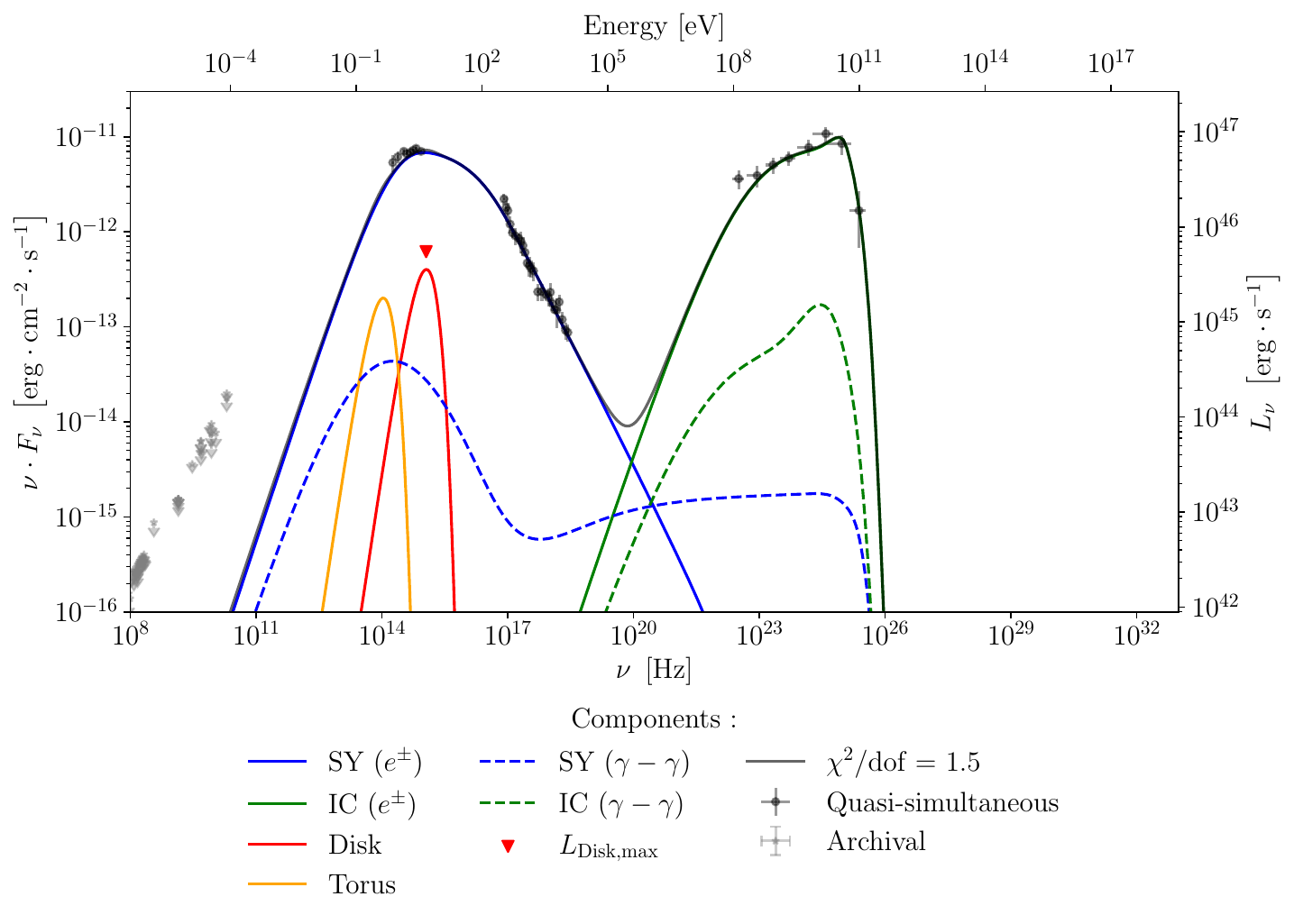}
    \caption{Spectral energy distribution from the purely leptonic model of \target~ in the observer frame. The dotted-dashed curves represent synchrotron and inverse Compton from pair production due to gamma-gamma absorption. The black solid squares represent the contemporaneous MWL data used for the modeling. Archival data are shown by gray solid dots. The red downward triangle represents the accretion disk limit from \citep{Ghisellini_2012}.}
    \label{fig: purely leptonic SED}
\end{figure*}

In our initial attempt to model the SED of \target, we considered only a population of relativistic electrons. The simulated SED is shown in Fig.~\ref{fig: purely leptonic SED}, and the fitted parameters can be found in Table \ref{tab: parameters for models} labeled L. In this figure, the labels SY and IC stand respectively for synchrotron and inverse Compton, from either the leptonic population ($e^\pm$) or from $\gamma-\gamma$ pair productions. In this scenario, the data points until the X-ray are explained by synchrotron emission only. The GeV gamma-ray data are partly explained by synchrotron self-Compton (SSC). The best solution found in that scenario is close to equipartition with $u'_{\rm e} / u'_{\rm b} \sim 3.9$, where $u'_{\rm e}$ is the electron energy density and $u'_{\rm b}$ the magnetic energy density. It should be noted that we did not take into account the presence of cold protons in this model. Here, we derived a relatively high value for $\gamma_{\rm e, min}$, this can be due to the prior acceleration (injection) of a truncated power-law from a region closer to the black hole. The cooling process might also be inefficient for lower energies, leading to the development of the low energy tail outside the zone model here \citep{Katarzynski_2006}, explaining the radio counterpart. In fact, the extended jet is expected to contribute to the integrated radio flux where those low energy electrons will cool down \citep{Plavin_2022}. Such a model is beyond the scope of this paper. Specific interactions between the electron and proton populations can also explain high $\gamma_{\rm e,min}$ values \citep{Zech_2021}.

To account for the limit proposed by \cite{Ghisellini_2012}, we included the black body emission of the accretion disk. The final disk luminosity obtained is $L_{\rm disk} = 4.8 \times 10^{45}~\rm{erg} \cdot \rm{s}^{-1}$, which explains the peaky feature observed at the highest $\gamma$-ray energies with external Compton interactions. In this model, the blob is located at a dissipation radius of $R_{\rm diss} / R_{\rm BLR} = 1.7$, indicating that the influence of the BLR radiation field is still significant. The dust torus black body is also present but subdominant. 

From the accretion disk parameters obtained, we can derive a black hole mass of $\sim 10^{10}M_\odot$ and an associated Eddington luminosity of $\sim 3 \times 10^{48}~\rm{erg} \cdot \rm{s}^{-1}$. As suggested in \cite{Sbarrato_2012}, we use the accretion regime $\eta = L_{\rm BLR} / L_{\rm Edd}$ as physical criteria to distinguish BL Lacertae (BL Lacs) from FSRQs. We find an accretion regime of $\eta \sim 2 \times 10^{-4}$, close to the boundary that separates FSRQs from BL Lacs \citep[$\eta = 5 \times 10^{-4}$,][]{Sbarrato_2012}. Finally, the shortest timescale variability $\tau_{\rm var}$ derived from this model is $3.7~\rm{days}$ in the observer's frame, close to the derived value from optical variability. 

Given the X-ray spectrum, we further tested the possibility to reproduce the break in the X-ray band with the purely leptonic model, but we were not able to find an acceptable solution ($\chi^2_{\rm d.o.f.} \leq 2$, by lowering $\gamma_{\rm e, min}$ for example).
This suggests that, if the break is present, it may be interpreted as the presence of additional processes. As we discuss in the following sections, a hadronic component is capable to successfully account for the broken spectral shape observed in the X-ray band as well as the putative neutrino emission.

\subsection{Lepto-hadronic solution}
\label{subsec: Lepto-hadronic solution}

\begin{figure*}
    \centering
    \includegraphics[width = \textwidth]{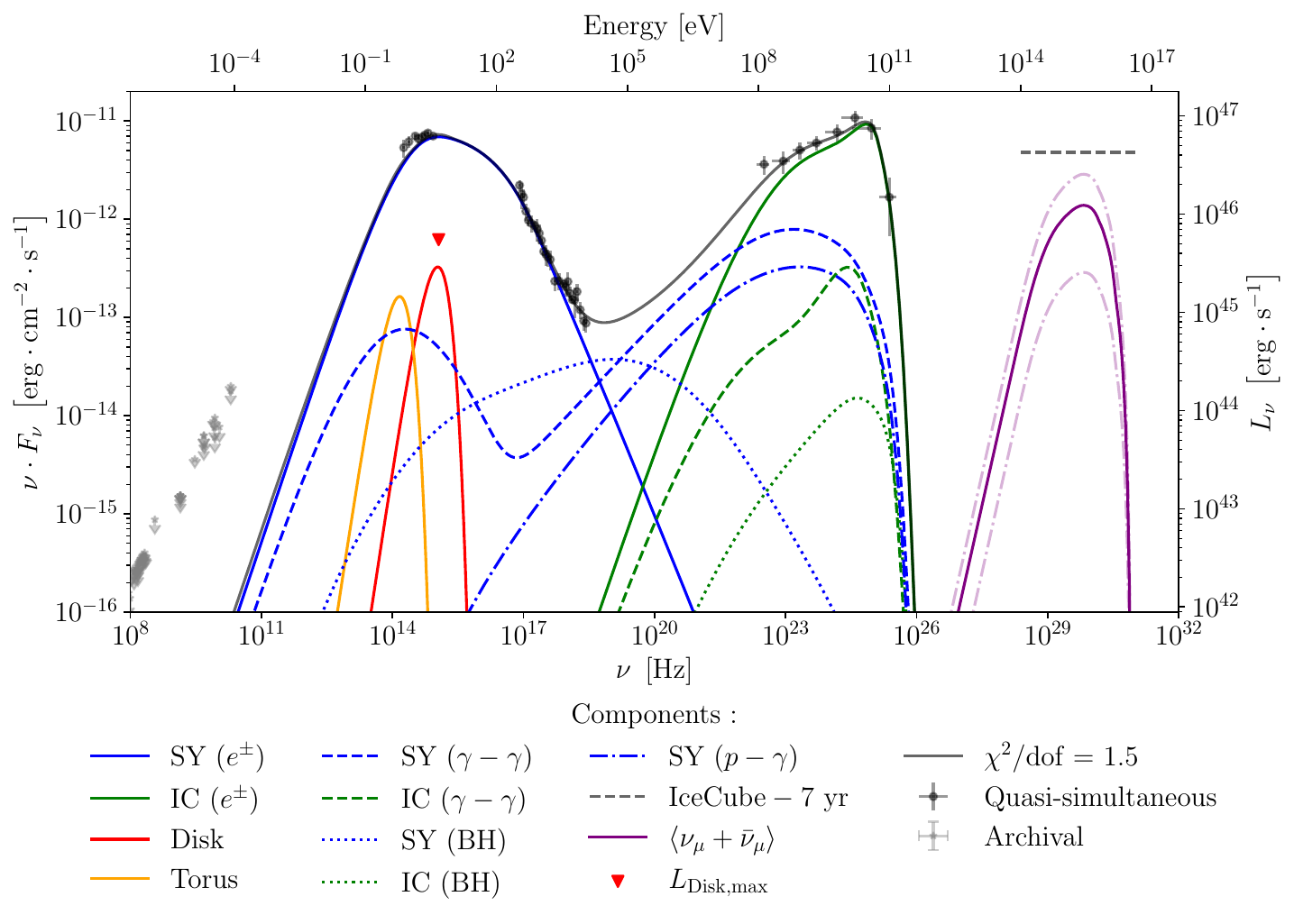}
    \caption{Same as Fig.~\ref{fig: purely leptonic SED} with the lepto-hadronic solution of \target~in the observer frame. Dashed curves represent hadronic processes (BH (Bethe-Heitler) and photo-pion productions). The dotted-dashed blue curves represent the obtained muon neutrino flux with associated uncertainties from Poisson statistic assuming a $3\sigma$ levels.}
    \label{fig: lepto hadronic SED 1}
\end{figure*}

We use the parameters found in the purely leptonic solution as starting point for the parameter space research of the lepto-hadronic scenario. The final parameters are displayed in the Tab.~\ref{tab: parameters for models}, labeled as LH, and the SED is displayed on the Fig.~\ref{fig: lepto hadronic SED 1}. Here BH stands from Bethe-Heitler reaction, while $p-\gamma$ stands for photo-pion production.
Protons are injected with a simple power-law index of $2.0$ and the hadronic processes remain globally sub-dominant, except in the X-ray and MeV bands. Within the lepto-hadronic solution, we find values that are in agreement with those found for the purely leptonic model with an accretion regime of $\eta \sim 2 \times 10^{-4}$, and an Eddington luminosity of $\sim 3 \times 10^{48}~\rm{erg} \cdot \rm{s}^{-1}$. Similarly, the shortest timescale variability derived here, $\tau_{\rm var} \sim 3.3~\rm{days}$, is consistent with the value derived from optical variability. The cascade component accounts for the X-ray flux, particularly at the highest energy.

The higher energy MWL peak is mainly explained by SSC and EC interactions with the BLR, consistently with the pure leptonic solution. Similarly, we find a dissipation radius close to $1.6~R_{\rm BLR}$. The cascade component accounts for the hard X-ray data, leading to a steeper index for the electron broken power-law. Indeed, significant contributions from $\gamma-\gamma$ pair production in the GeV energy band can also be observed, which are more pronounced than the ones observed in the leptonic model due to the additional photon fields from mesons disintegration. Although the model finds $u'_{\rm p} / u'_{\rm b} \sim 10^{3}$ (where $u'_{\rm p}$ is the proton energy density), far from the equipartition, the proton luminosity remains below the Eddington limit. 

The model presented here may be considered a viable solution for the efficient neutrino production case found by our modeling. As the X-ray data provide upper-limits for the cascade component, a broad range of solutions can be obtained with less energetic protons and, hence, lower predicted neutrino fluxes. 

\begin{figure}
    \centering
    \includegraphics[width = 0.5\textwidth]{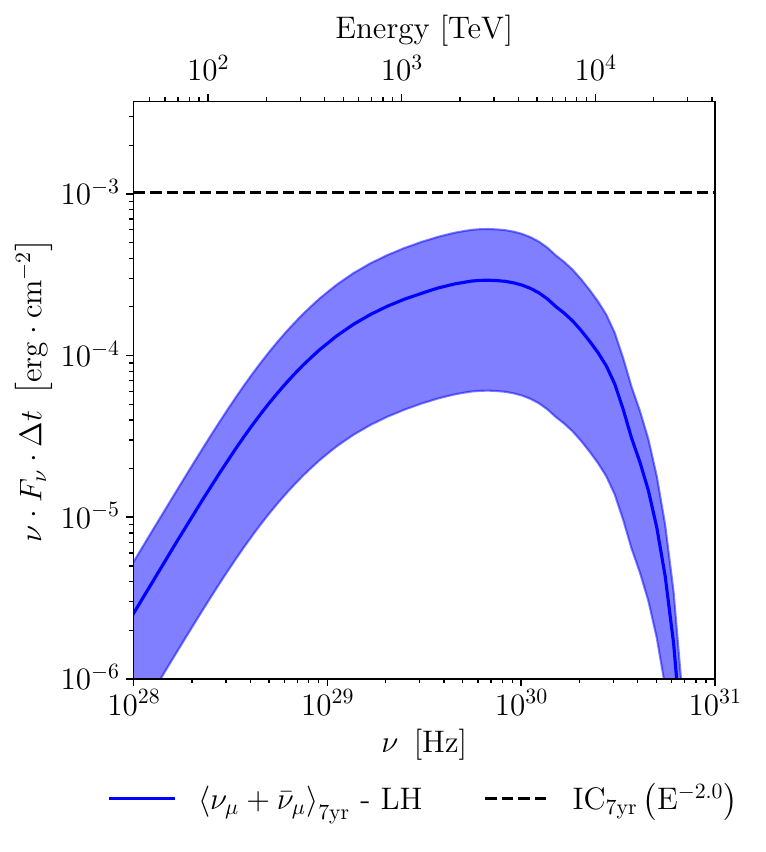}
    \caption{Neutrino flux scaled according to the livetime showed in \cite{Aartsen_2017b} derived for both mixed lepto-hadronic modeling of \target. The uncertainties are computed assuming Poisson statistics and $3\sigma$ levels. The dashed gray line represents the $7$-years IceCube sensitivity, assuming $E^{-2}$ and $\delta = -24\degree.06$.
    }
    \label{fig: neutrino flux}
\end{figure}

\bgroup
\def\arraystretch{1}
\begin{table}[]
    \centering
    \caption{Parameters used for the leptonic and the mixed lepto-hadronic models.}
    \label{tab: parameters for models}
    \begin{tabular}{l c c}
    \hline 
         & L & LH  \\
    \hline 
    $\delta_{\rm D}$ & $22.7$ & $22.5$ \\
    $R'_{\rm b}~\left[\rm{cm}\right]$ & $1.1 \times 10^{17}$ & $9.8 \times 10^{16}$\\
    $\tau_{\rm var}~\left[\rm{days}\right]$ & $3.7$ & $3.3$\\
    $B'~\left[\rm{G}\right]$ & $6.4 \times 10^{-2}$ & $ 8.3 \times 10^{-2}$ \\ 
    $u'_{\rm b}~\left[\rm{erg} \cdot \rm{cm}^{-3}\right]$ & $2.7 \times 10^{-4}$ & $3.1 \times 10^{-4}$ \\
    \hline 
    $\gamma'_{\rm e, min}$ & $10^{4}$ & $10^{4}$ \\
    $\gamma'_{\rm e, brk}$ & $1.1 \times 10^{5}$ & $1.3 \times 10^{5}$ \\
    $\gamma'_{\rm e, max}$ & $9.6 \times 10^{7}$ & $1.0 \times 10^{8}$ \\ 
    $p_{\rm e, 1}$ & $2.71$ & $2.73$ \\ 
    $p_{\rm e, 2}$ & $3.84$ & $4.26$ \\ 
    $u'_{\rm e}~\left[\rm{erg} \cdot \rm{cm}^{-3}\right]$ & $6.4 \times 10^{-4}$ & $6.3 \times 10^{-4}$\\
    $u'_{\rm e} / u'_{\rm b}$ & $3.9$ & $2.3$\\
    $L'_{\rm e}~\left[\rm{erg} \cdot \rm{s}^{-1}\right]$ & $1.2 \times 10^{42}$ & $1.0 \times 10^{42}$\\
    \hline 
    $\gamma'_{\rm p, min}$ & $-$ & $90$\\
    $\gamma'_{\rm p, max}$ & $-$ & $1.0 \times 10^{7}$\\ 
    $p_{\rm p}$ & $-$ & $2.0$\\
    $u'_{\rm p}~\left[\rm{erg} \cdot \rm{cm}^{-3}\right]$ &  $-$ & $1.5$\\
    $u'_{\rm p} / u'_{\rm b}$ & $-$ & $5.3 \times 10^3$\\
    $L'_{\rm p}~\left[\rm{erg} \cdot \rm{s}^{-1}\right]$ & $-$ & $1.0 \times 10^{45}$\\
    \hline 
    $L_{\rm disk}~\left[\rm{erg} \cdot \rm{s}^{-1}\right]$ & $4.8 \times 10^{45}$ & $3.9 \times 10^{45}$\\
    $T_{\rm disk}~\left[\rm{K}\right]$ & $1.4 \times 10^4$ & $1.3 \times 10^4$\\
    $T_{\rm torus}~\left[\rm{K}\right]$ & $1.3 \times 10^3$ & $1.3 \times 10^3$\\
    $R_{\rm diss} / R_{\rm BLR}$ & $1.7$ & $1.6$ \\
    \hline 
    $N_{\rm events}$ per year & $-$ & $0.68^{+2.32}_{-0.68}$ \\
    $N_{\rm events}$ (total)  & $-$ & $4.82^{+5.18}_{-3.82}$\\
    \hline
    $\chi^2/{\rm{d.o.f.}}$ & $1.5$ & $1.5$\\ 
    \hline
    \end{tabular}
\end{table}
\egroup

\section{Discussion}
\label{sec: Discussion}

Although still widely used, the historical classification of blazars based purely on observational characteristics, e.g. the optical spectrum and the location of the low-energy peak, has been put into question for a long time. Since the availability of large samples of blazars, thanks to $Fermi$-LAT observations, \citet{Ghisellini_transition:2011} has debated in favor of a more physical distinction for blazars based on the luminosity of the BLR measured in the Eddington units. Indeed, the BLR and Eddington luminosities are respectively related to the optical spectrum and the jet power (energy injected in the primary relativistic electrons). This can set a divide approximately where the disc transitions from a radiatively efficient to an inefficient regime. 

Prior to being included in the sample of PeVatron blazars and proposed as a high-energy neutrino emitter,  \target\ stood out in the literature due to its peculiarities. Historically, it has been classified as a BL Lac object due to the featureless optical spectrum and its high synchrotron peak, with $\nu_{\rm pk}^{\rm sy} \sim 10^{15}~\rm{Hz}$. 
\target\, was pinpointed as an exemplary blazar, it displays properties typical of “blue flat spectrum radio quasar” \citep[][a.k.a. “high-power high-synchrotron-peak blazars”]{Ghisellini_transition:2011,Padovani_2012,Ghisellini_2012}, i.e. high-emitting power sources that are intrinsically FSRQs where their broad emission lines are swamped by the jet synchrotron emission. For reference, this typology of blazars has also been  called “masquerading BL Lacs” \citep{Giommi_2013,Padovani_TXS_notBLLac:2019}.
In contrast to “true” high-frequency-peaked BL Lacs which have intrinsically poor radiation fields, these objects host powerful jets and radiatively efficient accretion. 

The study presented here allows us to provide conclusive evidence to the earlier speculations regarding the peculiar nature of \target. The SED modeling reveals a bright accretion disk with $L_{\rm disk} \sim 4 \times 10^{45}~\rm{erg} \cdot \rm{s}^{-1}$. The presence of external fields that are partly re-processed by the BLR naturally explains the peaky feature in the $\gamma$-ray band. 
Its accretion regime, i.e. the energy injected into these external fields, is of the order of $L_{\rm BLR} / L_{\rm Edd} \sim 2 \times 10^{-4}$, close to the values physically suggested for FSRQs \citep{Ghisellini_transition:2011}.
The relatively high accretion regime of \target~ is supported also by the ratio of the $\gamma$-ray luminosity $L_{\gamma}\left(0.1 - 100~\rm{GeV}\right)$ and the Eddington luminosity, $L_\gamma / L_{\rm Edd} \simeq 0.15$, highlighting that this source shares properties common to the FSRQ class \citep{Sbarrato_2012}.

The location of the emitting region, outside but close to the BLR radius, makes the BLR radiation influence on the blob still important while at the same time it limits the absorption of $\gamma$-rays, leading to a bright $\gamma$-ray luminosity. We find that in \target\ the combination of an efficient particle acceleration ($E_{\rm p, max} \simeq 10^{19}~\rm{eV}$) and efficient external radiation fields fosters the production of neutrinos, similarly to TXS~0506$+$056 and PKS~1424$+$240, suggested as promising candidate neutrino emitters \citep{Padovani_TXS_notBLLac:2019,Padovani_PKS_notBLLac:2022,Padovani_2022}.

Here, we propose a scenario of mixed lepto-hadronic models, where protons are injected with various energies. It is important to note that these solutions are conservative in terms of the total power injected into the hadrons. Emission from secondary particles in these models are mostly sub-dominant, although they can account for the hard X-ray and in the MeV bands. In fact, the broken spectral shape in the X-ray data suggests the presence of underlying processes. 

\section{Neutrinos from \target}
\label{subsection: 5BZB J0630-2406 as a neutrino emitter}

To estimate the number of neutrinos predicted in the lepto-hadronic scenario, we compute the energy flux for different flavors of neutrinos in the observer frame, and we consider the actual number of $\mu$-neutrinos according to oscillations. To account for the detection efficiency, we use tabulated effective areas from the point-source analysis of the IceCube detector in its final configuration with $86$ strings \citep[IC$86$,][]{Aartsen_2017b}, at the declination of \target~($\delta = -24^{\circ}.06$). We utilize the published effective areas presented in \cite{Aartsen_2017b}, for each previous configurations: IC$40$, IC$59$, IC$79$ and IC$86$. To determine the expected number of events, we integrate the flux over the energy range observed by IceCube ($300$ TeV - $1$ EeV) using the given effective area and over a period of $\sim7$ years, which is the full livetime of the observations used in \citet{Buson:2022}. For reference, we also display the IceCube $7$-year flux sensitivity derived at $\delta = -24\degree.06$ and assuming a neutrino spectrum of $E^{-2}$ \citep{Aartsen_2017}.

\begin{figure}
    \centering
    \includegraphics[width = \linewidth]{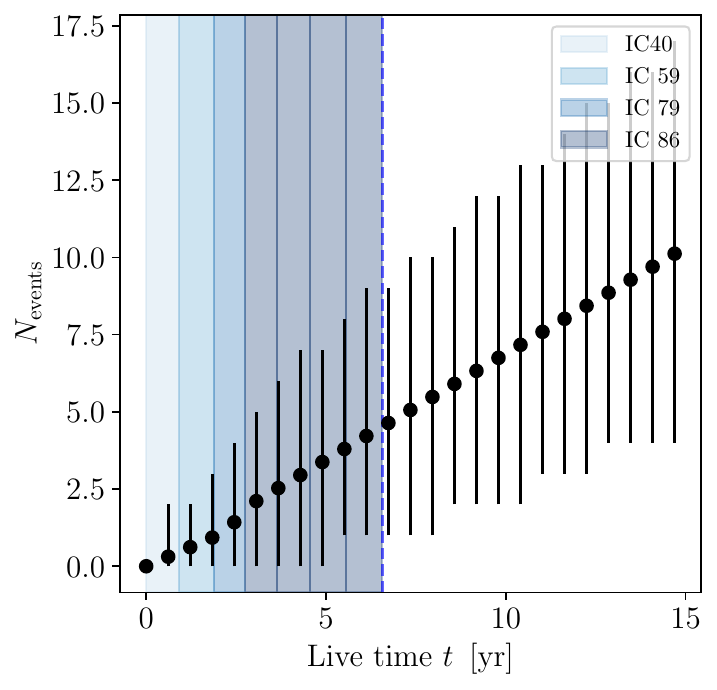}
    \caption{Evolution of the number of observed events by the IceCube detector in time according to various string configurations, and assuming a constant flux. The uncertainties are evaluated assuming Poisson statistics ($3\sigma$ levels). The blue dashed line represents the integrated livetime used in \citet{Buson:2022}.}
    \label{fig: N_events_p_value_LH1}
\end{figure}

The expected flux of $\mu$-neutrinos as observed on Earth is shown in Fig.~\ref{fig: neutrino flux} and is close to the IceCube sensitivity \cite[][estimated for a $\propto E^{-2}$ spectrum]{Aartsen_2017}. Our lepto-hadronic model predicts $N_{\rm events} = 4.82^{+5.18}_{-3.82}$ over a livetime period of $7$ years, as shown on Fig.~\ref{fig: N_events_p_value_LH1}. Fig. \ref{fig: N_events_p_value_LH1} shows the temporal evolution of the number of events $N_{\rm events}$ assuming a constant flux over the livetime of the IceCube detector, accounting for the different strings configurations, with $3\sigma$ uncertainties. 
Assuming low counting statistics, e.g. Poisson statistics, we can evaluate the probability of $N_{\rm events}$ detection against a null hypothesis of no detection, under the form of a $p$-value. Considering the first $7$-year livetime \cite{Aartsen_2017b}, we find a $p$-value of $0.03$, indicating a small tension with the null hypothesis of no detection. It should be noted that, in the most conservative scenario, our model still predicts a minimum of $N_{\rm min} = 4.82 - 3.82 = 1$ event, with a $p$-value higher than $0.05$. We find that \target~ contributes for $\leq 1\%$ to the IceCube diffuse muon neutrino flux \citep{IceCube_2022}. Similar values were obtained for TXS\,0506$+$056 \citep{Aartsen_2016, 2018Sci...361..147I}.

We can derive the baryonic loading in the blob frame, which represents the ratio between the proton luminosity and the $\gamma$-ray luminosity, denoted by $\xi = L'_{\rm p} / L'_{\rm \gamma}$. Based on our lepto-hadronic model, we find a loading factor of $\xi \sim 10^3$. This value appears notably high when compared to the typical order of magnitude, which is around $100$ as derived in \cite{Murase_2014}. Nevertheless, the value obtained here is consistent with the range derived in \cite{Palladino_2019}, as the intermediate BL Lac - FSRQ objects show a higher $\xi$, but still respecting the IceCube stacking limit \citep{Aartsen_2015}. Appendix \ref{sec: An ultra-high-energy cosmic ray event, position consistent with} discusses our findings in the context of literature studies that tackled \target~as an ultra-high-energy cosmic ray (UHECR) candidate. 

\section{Conclusions}
\label{sec: Conclusion}

\target\ has been recently proposed as a high-energy neutrino emitter based on the observation of a statistically significant spatial correlation with high-energy IceCube data. To address the plausibility of this association from the theoretical perspective, in this work we analyze simultaneous and quasi-simultaneous multi-wavelength data of the blazar, and model the emission of the source using a one-zone model in the context of both a purely leptonic and mixed lepto-hadronic scenario. \\
We summarize the main findings in the following.

\begin{itemize}
    \item Despite being formally classified as a high-synchroton-peaked BL Lac object, based on its featureless optical spectrum, the intrinsic nature of \target\ is that of a high-power FSRQ. It hosts a standard accretion disk and BLR; the optical emission lines elude direct observation, being the optical band swamped by the non-thermal continuum. The combination of efficient external radiation fields and enhanced particle acceleration efficiency offers ideal conditions for the production of neutrinos.
    \item The presence of a bright accretion disk is confirmed by a peaky feature in the $\gamma$-ray spectrum at the highest observable LAT energies. In FSRQ sources, a similar spectral shape may be observed and is naturally explained by external Compton reprocessing of the disk/BLR radiation by the jet. 
    \item The SED can be adequately modeled via both purely leptonic and mixed lepto-hadronic scenarios, suggesting that the hadronic component is sub-dominant except in the X-ray and in the MeV bands. Our results predict that future missions in the MeV band, such as AMEGO-x and ASTROGAM \citep{Caputo_2022, DeAngelis_2021}, hold the power to discriminate between these models.
    \item The analysis of the simultaneous \xmm\ and \nus~ spectra during a comparatively low state, provides evidence ($\gtrsim3\sigma$) of a break in the X-ray band. If the break was to be intrinsic to the object, a pure leptonic model faces challenges in reproducing it. On the other hand, the proposed lepto-hadronic model shows a turnover of the spectrum in the X-ray band, that marks the kick-in of the hadronic component contribution. Based on our theoretical modeling, the SED is overall leptonic-dominated. Therefore, observations in lower activity states, as the one studied here, may offer better chances to pinpoint the hadronic fingerprint.
    \item The relatively high accretion regime of 5BZB\,J0630-2406 is supported also by the ratio of the $\gamma$-ray luminosity $L_{\gamma}\left(0.1 - 100~\rm{GeV}\right)$ and the Eddington luminosity, $L_\gamma / L_{\rm Edd} \simeq 0.15$.
    As further reprove of the intrinsic FSRQ nature, the model reveals an efficient accretion regime of $2 \times 10^{-4}$. Besides, a relatively large fraction of the $\gamma$-ray luminosity ($\sim 15 \%$ of $L_{\rm Edd}$) is observed. Similarities can be found with the object TXS\,0506+056 \citep{Padovani_2019}, where the localization of the $\gamma$ emission region is thought to be, as in our case, on the edge of the BLR avoiding a significant $\gamma - \gamma$ absorption and an efficient neutrino production. 
    \item The neutrino emission predicted within the lepto-hadronic framework is at reach of the IceCube detector, close to the flux sensitivity. During the $7$-yr integration span of the data used in \citet{Buson:2022}, we expect to observe $N_{\rm events} = 4.82^{+5.18}_{-3.82}$ neutrinos. Assuming that the object maintains a constant neutrino rate over time, more high-energy neutrinos may be expected with increased instrument exposure. However, the uncertainties on the predicted numbers are large and long-term variability in the MWL light curve is clearly present at almost all observable frequencies. 
    \item The contribution of \target\ to the astrophysical neutrino diffuse flux is expected to be of the order of $\sim 1\%$.
\end{itemize}

Based on the theoretical predictions presented here, the PeVatron blazar \target\ is capable of producing neutrinos in the IceCube energy range and can plausibly contribute to the anisotropy observed in the distribution of IceCube events of the hotspot IC~J0630$-$2353  \citep[see Fig.~\ref{fig: p_map},][]{Buson:2022}.
\target\ is a high-power, radiatively efficient blazar. Other objects in the PeVatron blazar sample display similar characteristics, i.e. TXS\,0506$+$056, PKS\,1424$+$240 and 5BZB\,J0035$+$1515 \citep{Buson:2022, Buson_erratum:2022}.
At the current status it remains unclear whether this peculiar, relatively rare characteristic describes the persistent behavior of their engine and/or is linked to different environment properties, or may be tracing temporary physical changes, such as changes in the state of the accretion mode, as suggested for “changing-look blazars” \citep{Pena_2021}, or changes in the location of the dissipation region \citep{Ghisellini_redblue:2013}. Future investigation of the PeVatron blazar sample will enable us with a broader understanding of the neutrino/blazar physical relation.

\vspace{1cm}
\noindent
The authors thank the anonymous reviewer for the valuable comments and constructive feedback. 
The authors thank Andreas Zech, Anita Reimer, Markus Böttcher and Jörn Wilms for the fruitful discussions, and Weikang Zehng for providing the calibrated KAIT data.
GFDC thanks Xavier Rodrigues for the essential help with AM$^3$.
This work was supported by the European Research Council, ERC Starting grant \emph{MessMapp}, S.B. Principal Investigator, under contract no. 949555. This research has made use of the NASA/IPAC Extragalactic Database, which is funded by the National Aeronautics and Space Administration and operated by the California Institute of Technology (the NASA/IPAC Extragalactic Database (NED) is funded by the National Aeronautics and Space Administration and operated by the California Institute of Technology). 
Part of this work is based on observations obtained with the Samuel Oschin Telescope $48$-inch and the $60$-inch Telescope at the Palomar Observatory as part of the Zwicky Transient Facility project. ZTF is supported by the National Science Foundation under Grant No. \texttt{AST-2034437} and a collaboration including Caltech, IPAC, the Weizmann Institute for Science, the Oskar Klein Center at Stockholm University, the University of Maryland, Deutsches Elektronen-Synchrotron and Humboldt University, the TANGO Consortium of Taiwan, the University of Wisconsin at Milwaukee, Trinity College Dublin, Lawrence Livermore National Laboratories, and IN2P3, France. Operations are conducted by COO, IPAC, and UW. We acknowledge support by Institut Pascal at Université Paris-Saclay during the Paris-Saclay Astroparticle Symposium 2021 and 2022, with the support of the P2IO Laboratory of Excellence (program “Investissements d’avenir” ANR-11-IDEX-0003-01 Paris-Saclay and ANR-10-LABX-0038), the P2I axis of the Graduate School Physics of Université Paris-Saclay, as well as IJCLab, CEA, IPhT, IAS, OSUPS, the IN2P3 master projet UCMN, APPEC, and EuCAPT.

\bibliography{bib.bib}

\appendix

\section{An ultra-high-energy cosmic-ray event in the direction of \target}
\label{sec: An ultra-high-energy cosmic ray event, position consistent with}

An ultra-high-energy cosmic ray (UHECR) event has been observed in 2007 by the Pierre-Auger observatory, with an observed energy of $E_{\rm cr} = 60~\rm{EeV}$ and celestial coordinates $\rm{RA},~DEC = 105.9, -22.9$ \citep[$l,~b = -125.2, -7.7$,][]{Auger_anisotropies:2015}.
The blazar \target\ is located at an angular distance of $\theta = 7.6\degree$ from the UHECR event, and a previous study has considered it as a candidate counterpart \citep{Resconi:2017}.
Based on our theoretical modeling of \target, the maximum energy reached by the protons in the co-moving frame is $E'_{\rm p, max} \simeq 0.01~\rm{EeV}$. This translates to a maximum proton energy of $E_{\rm p, max} \simeq 0.2~\rm{EeV}$ in the observer frame, that is below the value of the ultra-high-energy observed. 
Cosmic rays propagate with Larmor radii $r_{\rm L} = 1.1~\rm{Mpc} \; E_{EeV} / (Z \times B_{\rm nG})$, where $E$ is the cosmic ray energy, $Z$ the atomic number and $B$ the magnetic field.
Using Hillas criterion and a minimal extragalactic magnetic field strength of $B = 1~\rm{nG}$, for the source of interest one can derive the Larmor radius  of $R_{\rm L, cr} \simeq 66~\rm{kpc}$, which represents the typical length scale of scattering and deflections. PeV-scale protons will scatter at the kpc scale, even for magnetic fields as weak as nG. In the vicinity of the source, the magnetic field may be much stronger, in which case the scattering/deflection length is even shorter. Consequently, at PeV-EeV energies, protons from cosmological distances are expected to be completely isotropized. It is therefore not possible to trace back the UHCRs to sources such as the one of interest here. 

\section{Parameter space research}
\label{sec: Parameter space research}

\begin{figure}
    \centering
    \includegraphics[width = 0.35\textwidth]{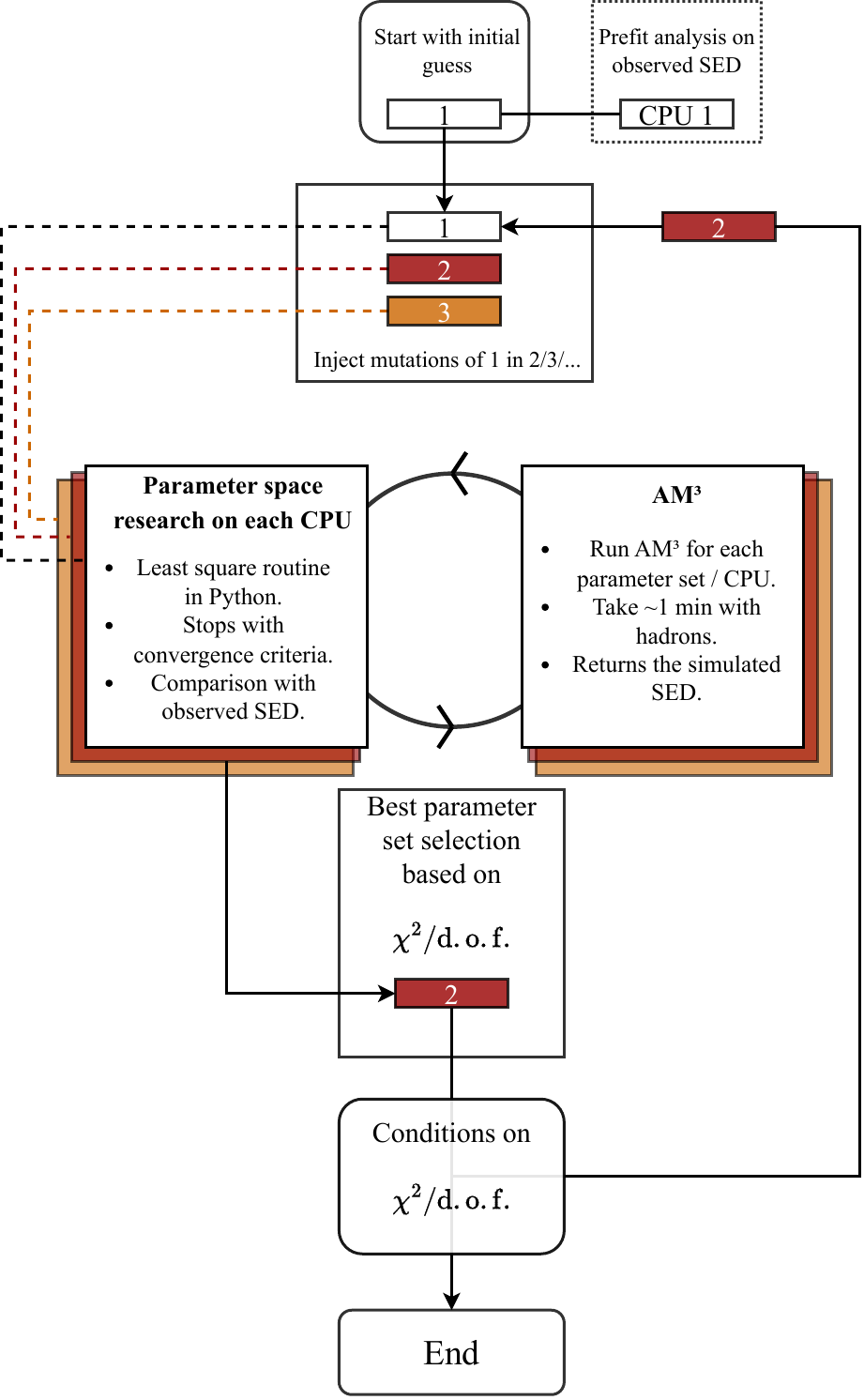}
    \caption{Flow chart of the parameter space research algorithm used in this study. A box labeled by a number represent a set of parameters for a given CPU.}
    \label{fig: optimization algorithm}
\end{figure}

\begin{figure}
    \centering
    \includegraphics[width = 0.5\textwidth]{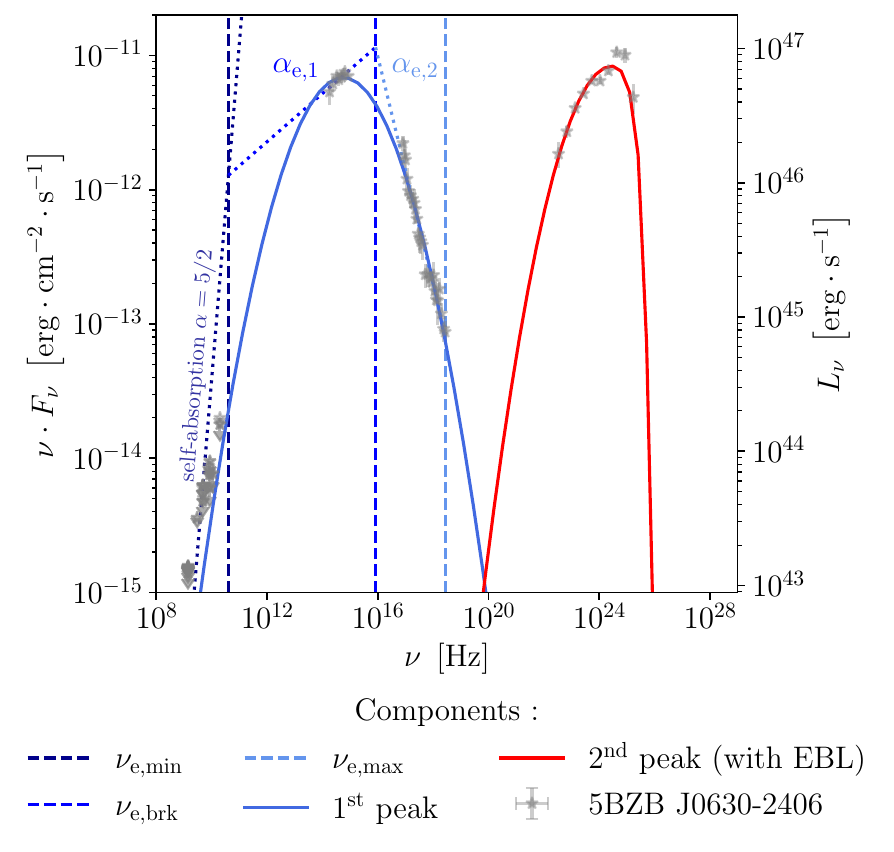}
    \caption{Pre-analyzed observed SED showing the first peak, assumed to be due to synchrotron emission from relativistic electrons, and the second peak. Assuming a Doppler factor $\delta_{\rm D}$, a magnetic field strength $B'$ (and the source redshift $z$), one can derive characteristics of the electron distribution, including the minimum, break, maximum energies and both indexes are indicated in the legend.}
    \label{fig: prefit SED}
\end{figure}

To explore the parameter space for a given scenario, we developed a comprehensive approach based on minimizing the residuals between the simulated and observed SEDs. This minimization method optimizes the goodness of fit by evaluating the reduced chi-squared ($\chi^2/{\rm d.o.f.}$) value. The flow chart of the optimization algorithm is illustrated in Fig.~\ref{fig: optimization algorithm}. \\
The first step involves a pre-fit of the observed SED (similar to \cite{Ghisellini_2012}). Assuming that the first bump-like feature arises from synchrotron emission by electrons, we can determine the values of $\gamma'_{\rm e, min}$, $\gamma'_{\rm e, break}$, and $\gamma'_{\rm e, max}$. Linear regression on data points within a given energy band enables us to derive spectral indices in both branches of the broken power-law distribution ($\alpha_{\rm e, 1}$ and, $\alpha_{\rm e, 2}$ respectively). The result of the pre-fit analysis is shown in the Fig.~\ref{fig: prefit SED}.\\

Assuming a broken power-law and synchrotron cooling, we find respectively $p_{\rm e, 1} = 2.65$ and $p_{\rm e, 2} = 3.69$. Furthermore, we can derive the energies in the jet frame as follows,

\begin{align}
    \gamma'_{\rm e, min} &= A_{\rm e, min} \cdot B'^{-1/2} / \delta_{\rm D} \cdot \left( 1 + z\right) \,, \\
    \gamma'_{\rm e, brk} &= A_{\rm e, brk} \cdot B'^{-1/2} / \delta_{\rm D} \cdot \left( 1 + z\right) \,, \\
    \gamma'_{\rm e, max} &= A_{\rm e, max} \cdot B'^{-1/2} / \delta_{\rm D} \cdot \left( 1 + z\right) \,,
\end{align}
where we have $A_{\rm e, min} = 1.7 \times 10^2$, $A_{\rm e, brk} = 4.5 \times 10^{4}$ and $A_{\rm e, max} = 8 \times 10^{5}$, the associated Lorentz factor obtained for values of Doppler factor $\delta_{\rm D} = 25$, magnetic field strength $B' = 0.1~\rm{G}$ and redshift $z = 1.239$. We also note that from the fit, we found $\nu_{\rm e, min} = 1.1 \times 10^{11}~\rm{Hz}$, $\nu_{\rm e, brk} = 8 \times 10^{15}~\rm{Hz}$ and $\nu_{\rm e, max} = 2.72 \times 10^{18}~\rm{Hz}$. \\

Concerning $\gamma'_{\rm e, min}$, we assume here that the radio flux corresponds to the integrated synchrotron emission along the jet, in the synchrotron self-absorption regime (assuming a spectral index $\alpha = 5/2$). \\
Finally, we used a log-parabola fit on both peaks to estimate the bolometric luminosity $L_{\rm bol} \sim 7.3 \times 10^{47}~\rm{erg} \cdot \rm{s}^{-1}$. This implies that the injected energy in the electrons should not exceed $L_{\rm bol} / \Gamma_{\rm D}^2$. These derived quantities were used as the initial guess. \\

\end{document}